# Modulation property of flexural-gravity waves on a water surface covered by a compressed ice sheet

**A.V. Slunyaev**[1,2], **Y.A. Stepanyants**[3,4;*)]

[1]National Research University – Higher School of Economics, 25 Bolshaya Pechorskaya Street, Nizhny Novgorod, 603950, Russia;
[2]Institute of Applied Physics of the Russian Academy of Sciences, 46 Ulyanov Street, Box-120, Nizhny Novgorod, 603950, Russia;
[3]School of Mathematics, Physics and Computing, University of Southern Queensland, 487–535 West St., Toowoomba, QLD, 4350, Australia;
[4]Department of Applied Mathematics, Nizhny Novgorod State Technical University, n.a. R.E. Alekseev, 24 Minin Street, Nizhny Novgorod, 603950, Russia.

**Abstract**

We study the nonlinear modulation property of flexural-gravity waves on a water surface covered by a compressed ice sheet of given thickness and density in a basin of a constant depth. For weakly nonlinear perturbations, we derive the nonlinear Schrödinger equation and investigate the conditions when a quasi-sinusoidal wave becomes unstable with respect to amplitude modulation. The domains of instability are presented in the planes of governing physical parameters; the shapes of the domains exhibit fairly complicated patterns. It is shown that under certain conditions the modulational instability can develop from shorter groups and for fewer wave periods than in the situation of deep-water gravity waves on a free water surface. The modulational instability can occur at the conditions shallower than that known for the free water surface $kh = 1.363$, where $k$ is the wavenumber and $h$ is the water depth. Estimates of parameters of modulated waves are given for the typical physical conditions of an ice-covered sea.

*)Corresponding author: Yury.Stepanyants@usq.edu.au

# 1. Introduction

Water waves in oceans, lakes and other estuaries covered by ice sheets more and more attract the attention of researchers in recent years. This interest is caused by the exploration of polar regions, which are rich in mineral resources. Works on the ice-covered seas are required by the development of infrastructure which includes the construction of dwellings, research stations monitoring weather and climate change, laboratories, aerodromes, barns, etc. In many countries lakes and rivers are covered by ice in the winter period which provides conditions for transport across the ice field. This also makes investigations of properties of the ice-water system topical. In such ice-covered zones, dangerous extreme wave events are registered repeatedly, examples are given in (Liu and Mollo-Christensen, 1988; Marko, 2003; Collins et al., 2015). Another problem, which is similar to the ice-water system, is related to large floating artificial constructions (airdromes, platforms, artificial islands, long tankers). The dynamic properties of such constructions are close to the properties of elastic ice sheets, and their description is based on the combination of classical hydrodynamic equations with specific boundary conditions on the free surface which account for the elastic plate within the Kirchhoff–Love model (Forbes, 1986; Părău and Dias, 2002; Il'ichev, 2016; 2021) or a special Cosserat theory of hyperelastic shells (Plotnikov and Toland, 2011; Guyenne and Părău, 2014).

There is a vast volume of publications devoted to the linear properties of flexural-gravity waves (FGWs); it is impossible to list here all publications therefore, we mention only the most relevant monographs (Kheisin, 1967; Squire et al., 1996; Sahoo, 2012; Bukatov, 2017). Much fewer publications are devoted to the non-linear processes occurring in the ice-water system. The weakly nonlinear theory was developed by Marchenko and Shrira (1992) using the Hamiltonian formalism and taking into account linear contributors to the pressure related to the rigidity of the ice and to the stresses due to external loads; in particular, the nonlinear Schrödinger (NLS) equation for directional waves in infinite depth was derived. Weakly non-linear modulated waves were studied by (Părău and Dias, 2002; Guyenne and Părău, 2014) within the framework of the NLS equation but neither the ice compression nor the ice-plate inertia was considered. In the paper by Il'ichev (2016) a modulated solitary wave of arbitrary amplitude in the form of a "bright" soliton was obtained by taking into account both these effects. In the recent publication, Il'ichev (2021) considered a strongly nonlinear envelope solitary wave within the framework of the primitive Euler equation for the particular carrier wavelength that corresponds to the minimum of the phase speed. Then, a similar solution in the form of NLS soliton was derived within the weakly nonlinear theory in finite-depth water and it was shown that both solutions are close to each other for a water basin of moderate depth. However, to the best of our knowledge, the general analysis of the modulational instability for the basic governing parameters was not carried out so far.

Below we fill this gap and derive the NLS equation for weakly nonlinear perturbations for a fluid of a finite depth. Meanwhile, we point misprints in the classic works by Ablowitz and Segur (1979, 1981) and emphasize inconsistency of the theory developed in Liu and Mollo-Christensen (1988). Then, on the basis of the Lighthill criterion, we investigate the conditions when a quasi-sinusoidal wave becomes unstable with respect to the amplitude modulation, i.e. for different relations of parameters, we derive the conditions when a small-amplitude modulation increases with time and becomes deeper due to the growth of side-bands in the spectrum. This allows us to determine zones on the plane of parameters where bright or dark envelope solitons can exist. As well known, such formation plays an important role in the oceanic wave dynamics (see, for example, (Kharif et al., 2009; Osborne, 2010)). In the recent decade, bright and dark solitons, as well as breathers were successfully reproduced in a series of laboratory experiments in hydrodynamic flumes with the open water surface (Chabchoub et al, 2011, 2012, 2013; Slunyaev et al, 2013). Long-lived envelope solitons embedded into fields of strongly irregular waves were found in numerical and laboratory simulations as well (see (Slunyaev, 2021) and numerous references therein). An observation of a giant nonlinear wave packet on the surface of the oceans was reported recently by Onorato et al. (2021). According to the numerical and laboratory simulations of hydrodynamic envelope solitons in open water (Slunyaev, 2009, Slunyaev et al., 2013), they are rather well described by soliton solutions of the weakly nonlinear NLS equation up to surprisingly big steepness, $ka_0$ ~ 0.1–0.2. Local wave breaking begins when $ka_0$ ~ 0.3. The values $ka_0$ ~ 0.05–0.1 are frequently considered as the characteristic steepness of nonlinear wind waves in the open seas; therefore, the NLS equation often serves as a reasonable first-order approximation model. Similar wave phenomena can occur with flexural-gravity waves in ice-covered ocean. Therefore, it is important to predict theoretically which amplitudes, widths, and speeds solitons can have. At which combination of ice-water parameters they can emerge. Our paper partially illuminates these issues.

The paper is organized as follows. In Section 2, we consider a model of flexural-gravity waves in the ocean covered by a compressed ice. In Section 3, we derive the NLS equation, and in Sections 4 and 5 we present the analysis of linear and nonlinear properties of the flexural-gravity waves, respectively. In the Discussion, we summarize the results obtained and present estimates for the typical parameters of modulated waves in an ice-covered sea.

## 2. A model of surface waves in the presence of ice cover

Let us consider plane waves which propagate along the horizontal $x$-axis with the $z$-axis directed upward. We assume that the fluid is ideal and irrotational, so that the velocity potential $\varphi(x, z, t)$ can be introduced, $\mathbf{v} = \nabla\varphi$, where $\mathbf{v}$ is the two-dimensional vector of fluid velocity, and $\nabla$ is the Hamiltonian operator in the $(x, z)$-plane. Then, we obtain from the equation of mass conservation for the incompressible fluid the Laplace equation in the domain occupied by the fluid:

$$\Delta\varphi = 0, \quad -h \leq z \leq \eta, \tag{1}$$

where $\eta(x, t)$ is the water surface displacement beneath the thin ice plate, and $\Delta \equiv \nabla^2 = \partial^2/\partial x^2 + \partial^2/\partial z^2$ stands for the Laplacian operator. The water rest level corresponds to the horizon, $z = 0$, while the flat bottom is at $z = -h$, where $h$ denotes the constant water depth. The non-leaking bottom boundary condition requires that

$$\frac{\partial\varphi}{\partial z} = 0, \quad z = -h. \tag{2}$$

On the upper boundary, $z = \eta$, we set the traditional kinematic condition which reflects the equality of two definitions of the vertical velocity component at $z = \eta$:

$$\frac{\partial\varphi}{\partial z} = \frac{\partial\eta}{\partial t} + \frac{\partial\varphi}{\partial x}\frac{\partial\eta}{\partial x}. \tag{3}$$

The other, the dynamic boundary condition, is the Bernoulli integral on the water surface $z = \eta$ where the role of the external pressure in the right-hand side plays a pressure produced by the bended elastic ice plate (see, for example, (Forbes, 1986; Squire et al. 1996; Sahoo 2012; Stepanyants and Sturova 2021)):

$$\frac{\partial\varphi}{\partial t} + g\eta + \frac{1}{2}\left[\left(\frac{\partial\varphi}{\partial x}\right)^2 + \left(\frac{\partial\varphi}{\partial z}\right)^2\right] = -\frac{1}{\rho}\left[D\frac{\partial^2}{\partial x^2}K(\eta) + QK(\eta) + M\frac{\partial^2\eta}{\partial t^2}\right], \tag{4}$$

where $\rho$ is the water density, $D = Ed^3/[12(1 - \nu^2)]$ is the coefficient of ice rigidity/elasticity, $E$ is the Young modulus of elastic plate, $Q$ is the coefficient of longitudinal stress ($Q > 0$ corresponds to the compression, and $Q < 0$ – to the stretching), and $M = \rho_1 d$. Other parameters are: $\nu$ is the Poisson ratio, $\rho_1$ is the ice density, $d$ is the thickness of the ice plate, and $g$ is the acceleration due to gravity. The function $K(\eta)$ describes the ice-plate curvature caused by the plate deflection; according to (Forbes, 1986) and (Il'ichev, 2016; 2021),

$$K(\eta) = \frac{\eta_{xx}}{\left[1+(\eta_x)^2\right]^{3/2} - d\eta_{xx}/2}. \tag{5}$$

Lower indices here denote partial derivatives with respect to *x*. Another model for the ice-plate curvature was suggested by Plotnikov and Toland (2011) and Guyenne and Părău (2014)). In (Liu and Mollo-Christensen, 1988) the NLS equation was derived in the limit of infinite depth, and the term with the coefficient *d* in Eq. (5) was omitted. Having parameter *M* proportional to *d*, this leads to a different expression for the nonlinear coefficient of the evolution equation.

The first term in the square brackets on the right-hand side of Eq. (4) describes the elastic property of the ice plate; the second term represents a longitudinal stress or strain of the plate; the third term describes the inertial property of the ice plate. Hereafter the coefficients *M*, *Q*, *D* and *d* will be treated as independent physical parameters.

Below we consider quasi-monochromatic perturbations of the velocity potential and related ice-plate deflection and derive the nonlinear Schrödinger equation for weakly nonlinear waves. Then, on the basis of the derived equation, we analyze the stability of such waves with respect to small amplitude modulations.

## 3. The weakly nonlinear theory for a modulated wave

To derive the nonlinear equation for long modulations of flexural-gravity waves from the system of governing equations (1)–(4), we use the asymptotic method employed in Slunyaev (2005). We choose the carrier wavenumber *k* and assume that the wave amplitude is small such that $k\eta = O(\varepsilon)$ where $\varepsilon \ll 1$ is a small parameter. We also assume that the wave is quasi-monochromatic with the narrow spectrum around the peak value *k* and width $\Delta k$ so that the relative spectrum width $\Delta k / k = O(\varepsilon)$ is of the same order of smallness as the wave steepness $k\eta$. Let us expand the wave fields in series of wave harmonics which will be treated separately:

$$\eta(x,t) = \eta_0(x,t) + \frac{1}{2}\left[\eta_1(x,t)E + \eta_{-1}(x,t)E^{-1} + \eta_2(x,t)E^2 + \eta_{-2}(x,t)E^{-2} + ...\right], \tag{6}$$

$$\phi(x,z,t) = \phi_0(x,z,t) + \frac{1}{2}\left[\phi_1(x,z,t)E + \phi_{-1}(x,z,t)E^{-1} + \phi_2(x,z,t)E^2 + \phi_{-2}(x,z,t)E^{-2} + ...\right], \tag{7}$$

where $E(x,t) = \exp(i\omega t - ikx)$. The term $\varphi_0$ represents the induced mean flow, and $\eta_0$ describes the long-scale surface displacement. The following conditions $\varphi_{-n} = \varphi_n^*$, $\eta_{-n} = \eta_n^*$ provide real-valued solutions for $\eta$ and $\varphi$, where asterisk stands for the complex-conjugate function and $n = 1, 2, 3, \ldots$ .

Each harmonic is additionally decomposed in the asymptotic series on the small parameter $\varepsilon \ll 1$:

$$\eta_n(x,t) = \varepsilon \eta_{n0}(x_1, t_1, t_2) + \varepsilon^2 \eta_{n1}(x_1, t_1, t_2) + \varepsilon^3 \eta_{n2}(x_1, t_1, t_2) + \ldots, \quad (8)$$

$$\varphi_n(x,z,t) = \varepsilon \varphi_{n0}(x_1, z, t_1, t_2) + \varepsilon^2 \varphi_{n1}(x_1, z, t_1, t_2) + \varepsilon^3 \varphi_{n2}(x_1, z, t_1, t_2) + \ldots, \quad (9)$$

where fast and slow coordinates, and fast and multi-scale slow times are introduced, such that the operations of differentiation in space and time act as follows:

$$\frac{\partial}{\partial x} \Rightarrow \frac{\partial}{\partial x_0} + \varepsilon \frac{\partial}{\partial x_1}, \quad \frac{\partial}{\partial t} \Rightarrow \frac{\partial}{\partial t_0} + \varepsilon \frac{\partial}{\partial t_1} + \varepsilon^2 \frac{\partial}{\partial t_2}. \quad (10)$$

In Eq. (8) $\eta_{n0} = 0$ if $n \neq \pm 1$, thus the dominant term corresponds to the first harmonic.

Functions in Eqs. (8) and (9) depend on slow horizontal coordinate and times. The only function $E(x, t)$ depends on the fast variables:

$$E(x, t) = \exp(i\omega t_0 - i k x_0). \quad (11)$$

Relying on the small surface displacements, the potential and its derivatives can be expanded into the Taylor series in the vicinity of surface $z = 0$. Then, the surface boundary conditions (3) and (4) give the following relations near $z = 0$:

$$\sum_{j=0}^{\infty} \frac{\eta^j}{j!} \frac{\partial^{j+1} \varphi}{\partial z^{j+1}} = \frac{\partial \eta}{\partial t} + \frac{\partial \eta}{\partial x} \sum_{j=0}^{\infty} \frac{\eta^j}{j!} \frac{\partial^j}{\partial z^j} \frac{\partial \varphi}{\partial x}, \quad (12)$$

$$\sum_{j=0}^{\infty} \frac{\eta^j}{j!} \frac{\partial^j}{\partial z^j} \frac{\partial \varphi}{\partial t} + g\eta + \frac{1}{2}\left(\sum_{j=0}^{\infty} \frac{\eta^j}{j!} \frac{\partial^j}{\partial z^j} \frac{\partial \varphi}{\partial x}\right)^2 + \frac{1}{2}\left(\sum_{j=0}^{\infty} \frac{\eta^j}{j!} \frac{\partial^{j+1} \varphi}{\partial z^{j+1}}\right)^2 = -\left(\hat{D}\frac{\partial^2 K}{\partial x^2} + \hat{Q}K + \hat{M}\frac{\partial^2 \eta}{\partial t^2}\right). \quad (13)$$

Now we introduce the notations $\{\hat{D}, \hat{Q}, \hat{M}\} = \{D/\rho, Q/\rho, M/\rho\}$ for the sake of convenience. The nonlinear function $K(\eta)$ is also expanded in the Taylor series. For the NLS theory it is sufficient to consider the three first terms of the expansion:

$$K(\eta) \approx \eta_{xx} + \frac{1}{2}d(\eta_{xx})^2 + \frac{1}{4}d^2(\eta_{xx})^3 - \frac{3}{2}\eta_{xx}(\eta_x)^2. \quad (14)$$

Here, as in Eq. (5), lower indices denote partial derivatives with respect to $x$. The series (6)–(9) are substituted in Eqs. (12)–(14) with the use of differentiation as per Eqs. (10) and bearing in mind Eq. (11). Then the terms for the given harmonic number $n$ and given order of smallness $m$, $\varepsilon^m E^n$ are collected and treated consequently.

The solution of the Laplace equation (1) with the bottom boundary condition (2) can be found explicitly for the given amplitudes $A_{nm}$ specified at the water surface as follows:

$\varepsilon^1 E^0$: 
$$\varphi_{00} = A_{00}(x_1, t_1, t_2),$$

$\varepsilon^2 E^0$:
$$\varphi_{01} = A_{01}(x_1, t_1, t_2),$$

$\varepsilon^3 E^0$:
$$\varphi_{02} = A_{02}(x_1, t_1, t_2) - \frac{1}{2}(z+h)^2 \frac{\partial^2 A_{00}}{\partial x_1^2},$$

$\varepsilon^1 E^n$, $n \neq 0$:
$$\varphi_{n0} = A_{n0}(x_1, t_1, t_2) \frac{\cosh nk(z+h)}{\cosh nkh},$$

$\varepsilon^2 E^n$, $n \neq 0$:
$$\varphi_{n1} = A_{n1}(x_1, t_1, t_2) \frac{\cosh nk(z+h)}{\cosh nkh} + i(z+h) \frac{\sinh nk(z+h)}{\cosh nkh} \frac{\partial A_{n0}}{\partial x_1},$$

$\varepsilon^3 E^n$, $n \neq 0$:
$$\varphi_{n2} = \frac{\cosh nk(z+h)}{\cosh nkh} \left[ A_{n2}(x_1, t_1, t_2) - \frac{(z+h)^2}{2} \frac{\partial^2 A_{n0}}{\partial x_1^2} + i(z+h)\tanh nk(z+h) \frac{\partial A_{n1}}{\partial x_1} \right].$$

The amplitudes of the velocity potentials $A_{nm}(x_1, t_1, t_2)$ in these equations are functions of slow variables. The nonlinear surface boundary conditions (12)–(14) constrain the amplitudes; the corresponding equations for the amplitudes are considered below in each order of smallness with respect to $\varepsilon$.

In the order $\varepsilon^1 E^1$, the well-known dispersion relation for the FGWs is found as the compatibility condition:

$$\omega^2 = \frac{k\sigma(g - k^2 \hat{Q} + k^4 \hat{D})}{1 + \sigma k \hat{M}}, \qquad \sigma = \tanh(kh). \tag{15}$$

Simultaneously, we obtain the relation between the leading-order surface displacement $\eta_{10}$ and the leading order amplitude of the velocity potential $A_{10}$:

$$\eta_{10} = i\lambda A_{10}, \qquad \lambda = -\frac{k\sigma}{\omega}. \tag{16}$$

Other harmonics are absent in the order $O(\varepsilon)$.

In the order $O(\varepsilon^2)$, three harmonics exist. At $\varepsilon^2 E^1$, the advection equation appears:

$$\frac{\partial A_{10}}{\partial t_1} + V \frac{\partial A_{10}}{\partial x_1} = 0, \tag{17}$$

where 
$$V = \frac{\omega(\sigma + kh(1-\sigma^2))}{2\sigma k(1+\sigma k\hat{M})} - \sigma \frac{k^2\hat{Q} - 2k^4\hat{D}}{\omega(1+\sigma k\hat{M})} =$$
$$= \frac{-2\sigma^2 k\hat{M}(k^2\hat{Q} - 2k^4\hat{D}) + \sigma(g - 3k^2\hat{Q} + 5k^4\hat{D}) + kh(1-\sigma^2)(g - k^2\hat{Q} + k^4\hat{D})}{2\omega(1+\sigma k\hat{M})^2}. \tag{18}$$

It can be readily confirmed that this expression is the group velocity of FGWs, $V = d\omega/dk$.

In the order $\varepsilon^2 E^1$, the relation between the surface displacement and the velocity potential has the form:

$$\eta_{11} = i\lambda\left(A_{11} + ip\frac{\partial A_{10}}{\partial x_1}\right), \quad \text{where} \quad p = \frac{V}{\omega} + \frac{\omega^2\left(k\hat{M}(kh+\sigma)+kh\sigma\right) + 2k\sigma\left(k^2\hat{Q} - 2k^4\hat{D}\right)}{k\omega^2\left(1+\sigma k\hat{M}\right)}. \quad (19)$$

The relation between the long-scale surface displacement and the amplitude of the wave induced potential, which appears in the order $\varepsilon^2 E^0$, is:

$$\eta_{01} = \frac{V}{g}\frac{\partial A_{00}}{\partial x_1} - \frac{k^2}{4g}\left[\left(1-\sigma^2\right) + d\hat{Q}\frac{\sigma^2 k^4}{\omega^2}\right]|A_{10}|^2. \quad (20)$$

The second harmonic is determined in the order $\varepsilon^2 E^2$:

$$A_{21} = i\chi A_{10}^2,$$

$$\text{where} \quad \chi = \frac{k^2(1+\sigma^2)}{8\sigma\omega}\frac{3\omega^2\left(\sigma^2-1+2\sigma k\hat{M}\right)+6\sigma k^3\left(\hat{Q}-5k^2\hat{D}\right)-k^4\sigma^2 d\left(\hat{Q}-4k^2\hat{D}\right)}{P_1},$$

$$P_1 = \omega^2\left(\sigma + 3k\hat{M}\right) + 3k\left(k^2\hat{Q} - 5k^4\hat{D}\right). \quad (21)$$

The corresponding term of the surface displacement is:

$$\eta_{21} = \frac{k\sigma}{4\omega^2}\frac{-8i\omega^3 A_{21} + k^2\left[\omega^2(1-3\sigma^2) + k^4 d\sigma^2\left(\hat{Q} - 4k^2\hat{D}\right)\right]A_{10}^2}{P_2}, \quad P_2 = \omega^2(1+\sigma^2) - \sigma P_1. \quad (22)$$

The nonlinear induced flow, which is supposed to propagate with the group velocity $V$, is defined by the following equation, which appears in the order $\varepsilon^3 E^0$:

$$(gh - V^2)\frac{\partial A_{00}}{\partial x_1} = \Gamma|A_{10}|^2, \quad \Gamma = -\frac{k^2}{2}\left(V\frac{1-\sigma^2}{2} + \frac{g\sigma}{\omega} + dV\hat{Q}\frac{\sigma^2 k^4}{2\omega^2}\right). \quad (23)$$

In the order $\varepsilon^3 E^1$, we finally obtain the evolution equation for the carrier wave amplitude, which accounts for the effects of nonlinearity and dispersion:

$$i\frac{\partial A_{10}}{\partial t_2} + \beta\frac{\partial^2 A_{10}}{\partial x_1^2} + q_1|A_{10}|^2 A_{10} + q_2 A_{10}\frac{\partial A_{00}}{\partial x_1} = 0, \quad (24)$$

where $\beta = -\frac{1}{2}\frac{d^2\omega}{dk^2} =$

$$\frac{1}{8\omega^3\sigma^2 k^2(1+\sigma k\hat{M})^2}\left[\omega^4\left\{(kh)^2(1+2\sigma^2-3\sigma^4) - 2\sigma kh(1-\sigma^2) + \sigma^2 + 4\sigma k\hat{M}\left[kh(kh+\sigma)(1-\sigma^2)+\sigma^2\right]\right\}\right.$$

$$\left. + 4\sigma^2\omega^2 k\left\{2\sigma(k^2\hat{Q} - 4k^4\hat{D}) + kh(1-\sigma^2)(k^2\hat{Q} - 2k^4\hat{D}) + \sigma^2 k\hat{M}(k^2\hat{Q} - 6k^4\hat{D})\right\} + 4\sigma^4 k^2\left(k^2\hat{Q} - 2k^4\hat{D}\right)^2\right], \quad (25)$$

$$q_1 = \frac{k^2}{1+\sigma k\hat{M}}\left\{-\frac{\chi}{1+\sigma^2} + \frac{\chi(T_1 + kdT_2) + T_3 + kdT_4 + (kd)^2 T_5}{P_2}\right\}, \quad (26)$$

$$T_1 = -\frac{\omega^2}{2}(1-\sigma^2), \quad T_2 = -2k\sigma^2\left(k^2\hat{Q} - k^4\hat{D}\right),$$

$$T_3 = \frac{k}{16\omega^3 g\sigma} \times \left\{-8\omega^6(1-\sigma^2)^2 + k\sigma\omega^4\left[g(5+12\sigma^2+3\sigma^4) - 12g\sigma^3 k\hat{M} - 24k^4\hat{D}(1-\sigma^2)^2\right] + \right.$$

$$\left. + 3g\sigma^4 k^2\omega^2\left[-g + 3g\sigma k\hat{M} + 4k^4\hat{D}(5-3\sigma k\hat{M})\right] - 9g^3\sigma^5 k^3 + 36g^2\sigma^5 k^7\hat{D}\right\},$$

$$T_4 = \frac{\sigma^2 k^5}{2\omega}\left(\frac{\hat{Q}}{8(1+\sigma k\hat{M})}\left\{\frac{8(1-\sigma^2)}{g}\left[k^2\hat{Q} - (4+3\sigma k\hat{M})k^4\hat{D}\right] - 7 + 15\sigma^2 + (1+7\sigma^2)\sigma k\hat{M}\right\} + (1-2\sigma^2)k^2\hat{D}\right)$$

$$T_5 = \frac{\sigma^4 k^6}{8\omega^3}\frac{4+\sigma k\hat{M}}{1+\sigma k\hat{M}}\left[k^6\hat{D}^2 + k^2\hat{Q}\frac{(1-2\sigma k\hat{M})\hat{Q} - (5-\sigma k\hat{M})k^2\hat{D}}{4+\sigma k\hat{M}} - \frac{3}{4}g\frac{1-3\sigma k\hat{M}}{4+\sigma k\hat{M}}(\hat{Q}-k^2\hat{D})\right],$$

$$q_2 = \frac{k}{1+\sigma k\hat{M}}\left[1 + \frac{\omega V}{2g\sigma}(1-\sigma^2)\right]. \tag{27}$$

Seemingly, the term $T_3$ is given in a compact form, when the coefficient $\hat{Q}$ is eliminated using the dispersion relation (15). Note that in all interesting cases the ice is supposed to be thin compared to the wavelength, and then, the parameter $kd$ is small. The terms $T_2$, $T_4$ and $T_5$ can be neglected in Eq. (26) under the assumption that $kd \ll 1$.

Using Eq. (23), we reduce Eq. (24) to the closed form:

$$i\frac{\partial A_{10}}{\partial t_2} + \beta\frac{\partial^2 A_{10}}{\partial x_1^2} + \tilde{\alpha}|A_{10}|^2 A_{10} = 0, \tag{28}$$

$$\text{where } \tilde{\alpha} = q_1 + \frac{q_2 k^2}{2(V^2-gh)}\left(V\frac{1-\sigma^2}{2} + \frac{g\sigma}{\omega} + dV\hat{Q}\frac{\sigma^2 k^4}{2\omega^2}\right). \tag{29}$$

Finally, the NLS equation for the complex amplitude $B(x,t) = \eta_1 = \varepsilon\eta_{10} + O(\varepsilon^3)$ of the surface displacement can be obtained by combining Eqs. (17) and (28) and using the relation (16):

$$i\left(\frac{\partial B}{\partial t} + V\frac{\partial B}{\partial x}\right) + \beta\frac{\partial^2 B}{\partial x^2} + \alpha|B|^2 B = 0, \quad \alpha = \frac{\tilde{\alpha}}{\lambda^2}, \tag{30}$$

which is accurate to the order $O(\varepsilon^3)$. Here the real ice-sheet displacement and the velocity potential are specified by the relations:

$$\eta(x,t) = \text{Re}\left[B\exp(i\omega t - ikx)\right], \quad \varphi(x,z,t) = -\frac{\omega}{k\sigma}\text{Im}\left[B\exp(i\omega t - ikx)\right]\frac{\cosh k(z+h)}{\cosh kh}. \tag{31}$$

Explicit expressions for the main coefficients of the theory are given in the electronic supplement to the paper.

The coefficients of Eq. (30) can be significantly simplified in the limit of infinite depth, $kh \to \infty$. They are listed below with the subindex $\infty$:

$$\omega_\infty^2 = \frac{k\left(g - k^2\hat{Q} + k^4\hat{D}\right)}{1 + k\hat{M}}, \quad V_\infty = \frac{\omega_\infty^2 - 2k^3\left(\hat{Q} - 2k^2\hat{D}\right)}{2\omega_\infty k\left(1 + k\hat{M}\right)}, \qquad (32)$$

$$\lambda_\infty = -\frac{k}{\omega_\infty}, \quad q_{1,\infty} = \lambda_\infty^2 \alpha_\infty, \quad q_{2,\infty} = \frac{k}{1 + k\hat{M}},$$

$$\beta_\infty = -\frac{1}{2}\frac{d^2\omega_\infty}{dk^2} = \frac{\omega_\infty^4\left(1 + 4k\hat{M}\right) + 4\omega_\infty^2 k^3\left[k\hat{M}\left(\hat{Q} - 6k^2\hat{D}\right) + 2\left(\hat{Q} - 4k^2\hat{D}\right)\right] + 4k^6\left(\hat{Q} - 2k^2\hat{D}\right)^2}{8\omega_\infty^3 k^2\left(1 + k\hat{M}\right)^2},$$

$$\alpha_\infty = \frac{k^2}{32\omega_\infty\left(1 + k\hat{M}\right)\left[\omega_\infty^2\left(1 + 3k\hat{M}\right) + 3k^3\left(\hat{Q} - 5k^2\hat{D}\right)\right]}\left\{8\omega_\infty^4\left(2 + 3k\hat{M}\right) + \right.$$

$$+ k^3\omega_\infty^2\left[6\left(5\hat{Q} - 21k^2\hat{D}\right) + 18k\hat{M}\left(\hat{Q} - k^2\hat{D}\right) + 4kd\left(5\hat{Q} - 8k^2\hat{D}\right) - 3k^2d^2\left(\hat{Q} - k^2\hat{D}\right)\left(1 + 3k\hat{M}\right)\right] +$$

$$\left. + k^6\left(\hat{Q} - k^2\hat{D}\right)\left[18\left(\hat{Q} - 5k^2\hat{D}\right) - k^2d^2\left(\hat{Q} - 13k^2\hat{D}\right)\right]\right\}.$$

The nonlinear coefficient $\alpha_\infty$ can be further simplified assuming that $kd$ is small. Note that due to the factor $(gh - V^2)$ in the left-hand side of Eq. (23) which is usually very big whereas the expression in the right-hand side of this equation is finite, the wave induced mean flow proportional to $|B|^2$ vanishes in the limit of infinite depth; this is the standard situation in the classic NLS theory which results in $\tilde{\alpha}_\infty \approx q_{1,\infty}$. The derived NLS equation can be improved taking into account higher-order terms of the asymptotic expansions and by a more accurate description of the mean flow structure with the help of the approach suggested by Dysthe (1979), and Trulsen and Dysthe (1996).

## 4. Analysis of the dispersion relation of flexural-gravity waves

For the analysis of the dispersion relation (15) it is convenient to use the dimensionless variables introducing the scaled quantities as follows:

$$\Omega = \omega\sqrt[4]{\frac{D}{\rho g^3}}, \quad \kappa = k\sqrt[4]{\frac{D}{\rho g}}, \quad \tilde{Q} = \frac{Q}{\sqrt{\rho g D}}, \quad \tilde{M} = \frac{\rho_1}{\rho}d\sqrt[4]{\frac{\rho g}{D}}, \quad \tilde{H} = h\sqrt[4]{\frac{\rho g}{D}}. \qquad (33)$$

Then the dispersion relation (15) is represented in the form which contains the minimal number of dimensionless parameters, $\tilde{Q}, \tilde{M}$, and $\tilde{H}$:

$$\Omega(\kappa) = \sqrt{\frac{\sigma\kappa\left(1 - \kappa^2\tilde{Q} + \kappa^4\right)}{1 + \sigma\kappa\tilde{M}}}, \qquad \sigma = \tanh\left(\kappa\tilde{H}\right). \qquad (34)$$

For the estimates, let us choose the following realistic values of parameters that are traditionally used in works devoted to flexural-gravity waves in ice-covered oceans: $\rho_1 =$

922.5 kg/m³; $d = 1$ m; $E = 5 \cdot 10^9$ Pa; $\nu = 0.3$; $\rho = 1025$ kg/m³; $h = 100$ m; $g = 9.81$ m/s². The compression parameter $Q$ can vary in a wide range but not exceed the limiting value $Q_*$ when the buckling phenomenon begins. The threshold value of the compression parameter $Q_*$ can be found from the condition of appearing of a double root in the fourth-degree polynomial under the square root in Eq. (34); this gives $\tilde{Q}_* = 2$. With the chosen set of parameters, we obtain: $Q_* = 4.291 \cdot 10^6$ kg/s², $\tilde{M} = 0.062$, $\tilde{H} = 6.846$. The typical plot of the dispersion relation (34) is shown in Fig. 1 for different values of $\tilde{Q}$.

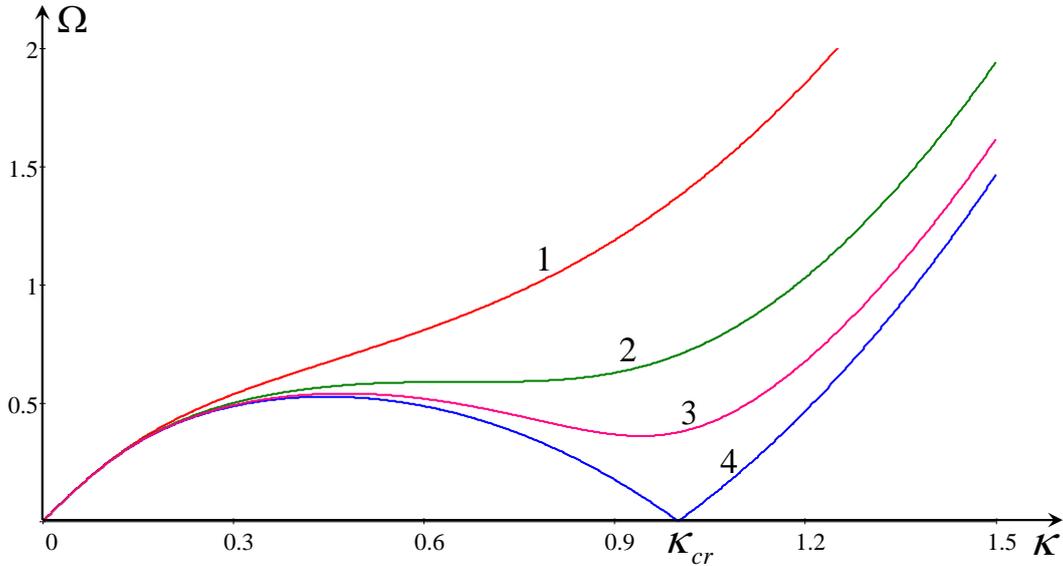

Fig. 1. (Color online.) The typical dispersion relation (34) for $\tilde{M} = 0.062$, $\tilde{H} = 6.846$ and several values of $\tilde{Q}$. Line 1 – $\tilde{Q} = 0$; line 2 – $\tilde{Q} = \tilde{Q}_0 \approx 1.475$; line 3 – $\tilde{Q} = 1.85$; line 4 – $\tilde{Q} = \tilde{Q}_* = 2$.

Let us start our analysis of the dispersion relation (34) with the case when there is no ice compression, $\tilde{Q} = 0$. Then, the gravity effect predominates over the elastic effect for small $\kappa < 1$, whereas for $\kappa > 1$, the elastic effect prevails. These effects are of a similar strength when $\kappa \approx 1$. The effect of a finite depth is important only for gravity waves when $\kappa < 1$ because in the typical situations the dimensionless water depth is relatively big, $\tilde{H} = 6.846$, so that $\sigma = \tanh(\kappa \tilde{H})$ is close to unity when $\kappa > 0.5$. The dimensionless depth can be smaller and its range of influence on the $\kappa$-axis becomes wider if the real water depth is $h < 10$ m.

The parameter $\tilde{M}$ which characterizes the inertial property of the ice plate usually is relatively small, $\tilde{M} < 0.1$, therefore, in many cases it can be neglected. Its influence becomes significant for a thick ice plate with $d > 5$ m. As one can see from Fig. 1, the most important influence on the dispersion curve is produced by the ice compression. Line 1 in the figure shows the dispersion relation with no compression, $\tilde{Q} = 0$; it is qualitatively similar to the dispersion curve of gravity-capillary waves with the monotonic increase of frequency with the wavenumber. When the compression increases, the dispersion curve bends down so that the local maximum and minimum appear on the curve when $\tilde{Q} > \tilde{Q}_0 \approx 1.475$. The maximum and minimum coalesce at $\tilde{Q} = \tilde{Q}_0$ (see line 2 in the figure). Further increase of the compression leads to the buckling phenomenon when $\tilde{Q} > \tilde{Q}_* = 2$; the dispersion relation becomes imaginary which implies that the destruction of the ice cover occurs around the critical point $\kappa_{cr}$. Graphics of the group velocities for the same values of $\tilde{Q}$ as in Fig. 1 are shown in Fig. 2.

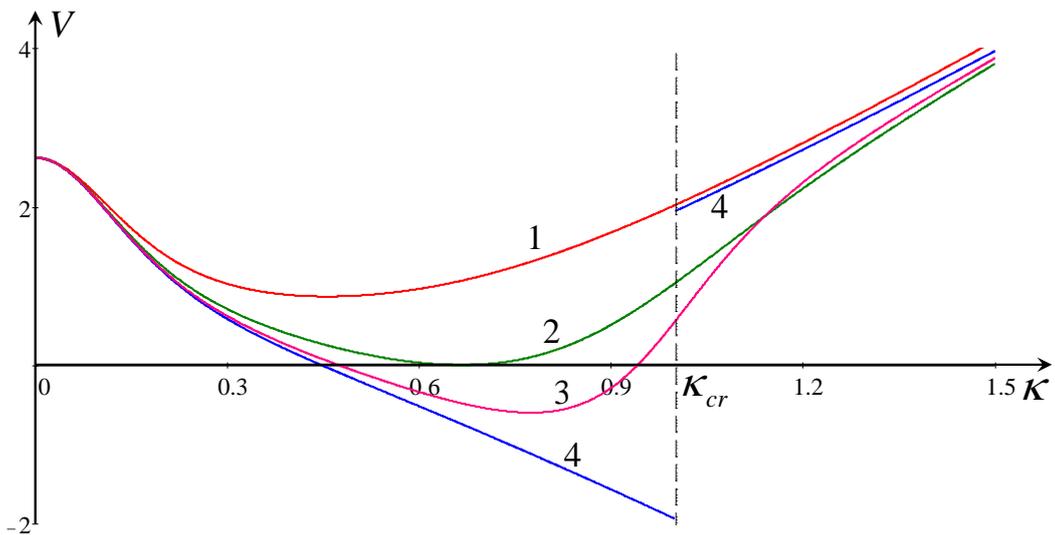

Fig. 2. (Color online.) The dependence of group velocity of FGWs on the wavenumber for $\tilde{M} = 0.062$, $\tilde{H} = 6.846$ and several values of $\tilde{Q}$. Line 1 – $\tilde{Q} = 0$; line 2 – $\tilde{Q} = \tilde{Q}_0 \approx 1.475$; line 3 – $\tilde{Q} = 1.85$; line 4 – $\tilde{Q} = \tilde{Q}_* = 2$. Dashed vertical line shows the critical wavenumber $\kappa_{cr}$ where line 4 splits into two branches, lower and upper.

As one can see from Fig. 2, the group velocity has a minimum at the intermediate wavenumbers. When the compression parameter increases, the point of minimum in the $\kappa$-space shifts toward $\kappa_{cr}$. The group velocity can be negative in a certain range of $\kappa$ when $\tilde{Q}_0 < \tilde{Q} < \tilde{Q}_*$. The slope of the dispersion curve at a certain point $\kappa$ determines the dispersion coefficient $\beta$ in the NLS equation (30), see Eq. (25).

## 5. The Benjamin–Feir instability of flexural-gravity waves

As well-known, waves described by the NLS equation (30) are affected by the modulational (alias Benjamin–Feir) instability, when the Lighthill criterion $\alpha\beta > 0$ is fulfilled (Lighthill, 1965; 1978; Ablowitz and Segur, 1981; Ostrovsky and Potapov, 1999). According to this criterion, a sufficiently long modulation of a quasi-monochromatic wave increases and becomes deeper when the nonlinear and dispersion coefficients in the NLS equation are of the same sign; otherwise, a small-amplitude modulation is not enhanced. The dispersion coefficient $\beta$ is strictly positive for free gravity water-waves in basins of constant depth without ice cover. The nonlinear coefficient $\alpha$ is positive in deep water and negative in shallow water due to the opposite effects of the nonlinear frequency shift and induced mean flow, when the coefficient $\Gamma$ in Eq. (23) is strictly negative. The nonlinear coefficient vanishes when $kh \approx 1.363$.

The effect of a surface tension on the modulation instability has been also studied; the diagram of the modulational instability of gravity-capillary waves can be found in the book by Ablowitz and Segur (1981) (see Fig. 4.15 there). In our notations, this problem formally corresponds to the following choice of parameters: $D = 0$, $M = 0$, $d = 0$, but $Q \neq 0$. In Fig. 3 we present a similar stability diagram for the dimensionless parameters of the water depth $kh$ and the longitudinal stress $k^2 \hat{Q}/g$. The stability diagram in the book by Ablowitz and Segur (1981) corresponds to the leftward part of our Fig. 3a where $Q \leq 0$, if Fig. 4.15 in the book is mirror-reflected with respect to the vertical axis. Shaded areas pertain to the domains of instability where $\alpha\beta > 0$. Unfortunately, using the expression for the nonlinear coefficient from (Liu and Mollo-Christensen, 1988), we failed to reproduce the result of Ablowitz and Segur (1981) in the limit $kh \to \infty$; therefore, we don't examine our results against the ones from (Liu and Mollo-Christensen, 1988). (Note that Ablowitz and Segur (1979) claimed that in their paper the derived equations are equivalent to the equations derived by Djordjevic and Redekopp (1977) "except for the correction of a misprint". In fact, they corrected the misprint in Eq. (2.12) of Djordjevic and Redekopp (1977) (where must be "2" rather than "$2\widetilde{T}$" in the numerator of the fraction on the right-hand side) but made another typo in Eq. (2.24d) (or Eq. (4.3.26) in the cited book), where must be $(3 - \sigma^2)$ rather than $(2 - \sigma^2)$. This misprint becomes obvious when the deep-water limit is considered – see Eq. (2.25) in Ablowitz and Segur (1979) or Eq. (4.3.27) in their book (1981).)

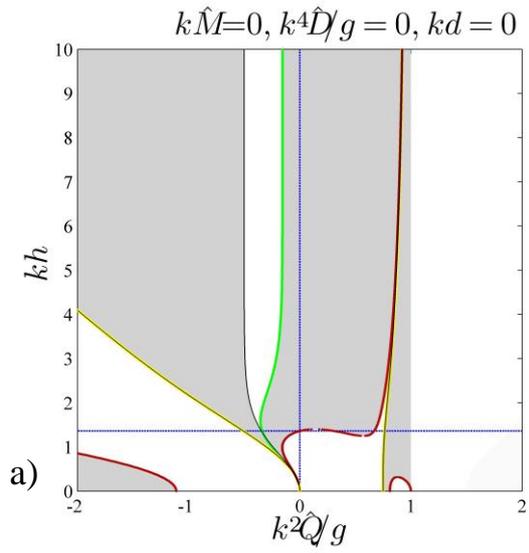
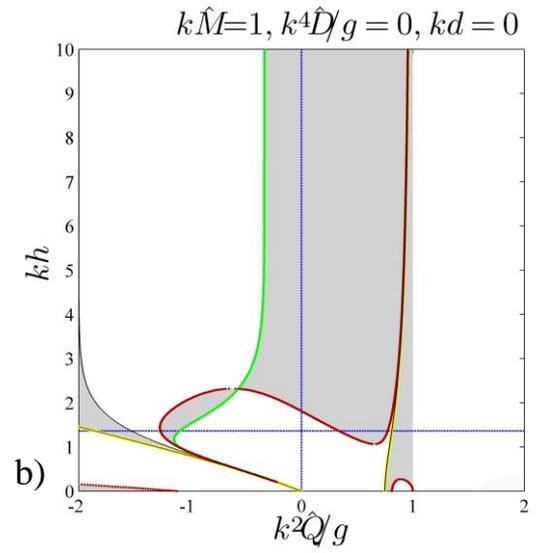
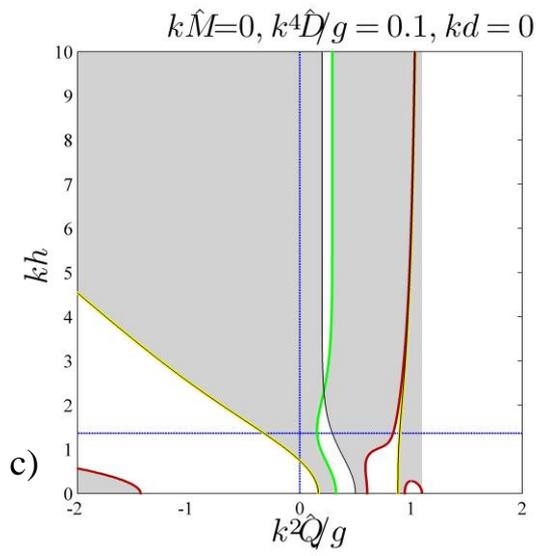
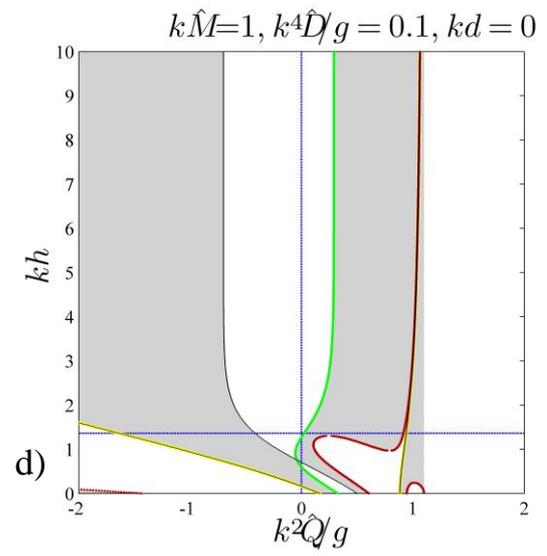
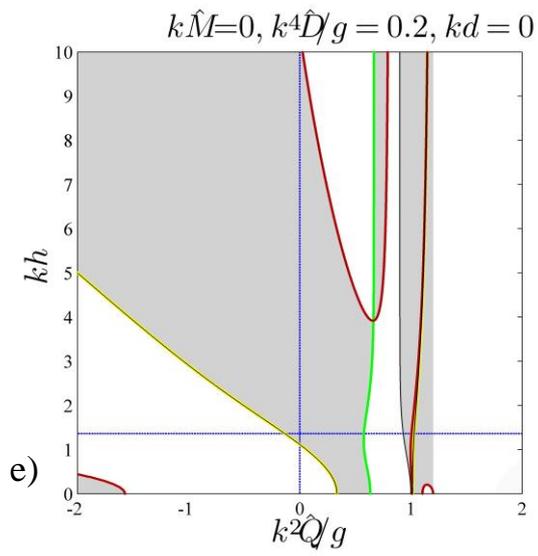
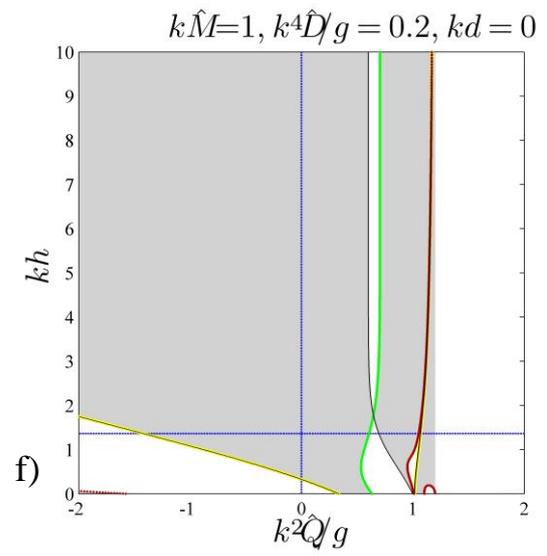

Fig. 3. (Color online.) Domains of the modulational instability (shaded) for $d = 0$. Frame (a): $\hat{M} = 0$, $\hat{D} = 0$; frame (b): $k\hat{M} = 1$, $\hat{D} = 0$; frame (c): $M = 0$, $k^4\hat{D}/g = 0.1$; frame (d): $k\hat{M} = 1$, $k^4\hat{D}/g = 0.1$; frame (e): $M = 0$, $k^4\hat{D}/g = 0.2$; frame (f); $k\hat{M} = 1$, $k^4\hat{D}/g = 0.2$. The curves where coefficient $\beta$ changes its sign are shown by the light-green color. The curves where the coefficient $\alpha$ turns to zero are shown by dark-red color. Black-yellow curves correspond to the condition of the resonance with the long-scale flow, while the black curves – to the resonance with short waves. Horizontal blue dotted lines designate the condition $kh = 1.363$.

Modifications of the stability diagram when $D$ and $M$ differ from zero are shown in other panels of Fig. 3. The rightmost boundaries on the diagrams are determined by the condition when the frequency (15) tends to zero, i.e.

$$g - k^2\hat{Q} + k^4\hat{D} = 0. \tag{35}$$

The domains of modulational instability in Fig. 3 look fairly complicated.

The general picture of the modulational instability is better represented by diagrams as functions of scaled water depth and rigidity parameter $D$ for the given values of $M$ and $Q$, which are presented in Figs. 4 and 5 in the way similar to Fig. 3. As expected, when the ice cover is absent ($\hat{Q} = 0$ in Figs. 3a and $\hat{D} = 0$ in Figs. 4a), water waves are unstable for $kh > 1.363$. The dotted horizontal blue lines in each frame in Figs. 3–5 depict the threshold $kh = 1.363$. The instability diagrams for several depth conditions are also plotted in Fig. 6.

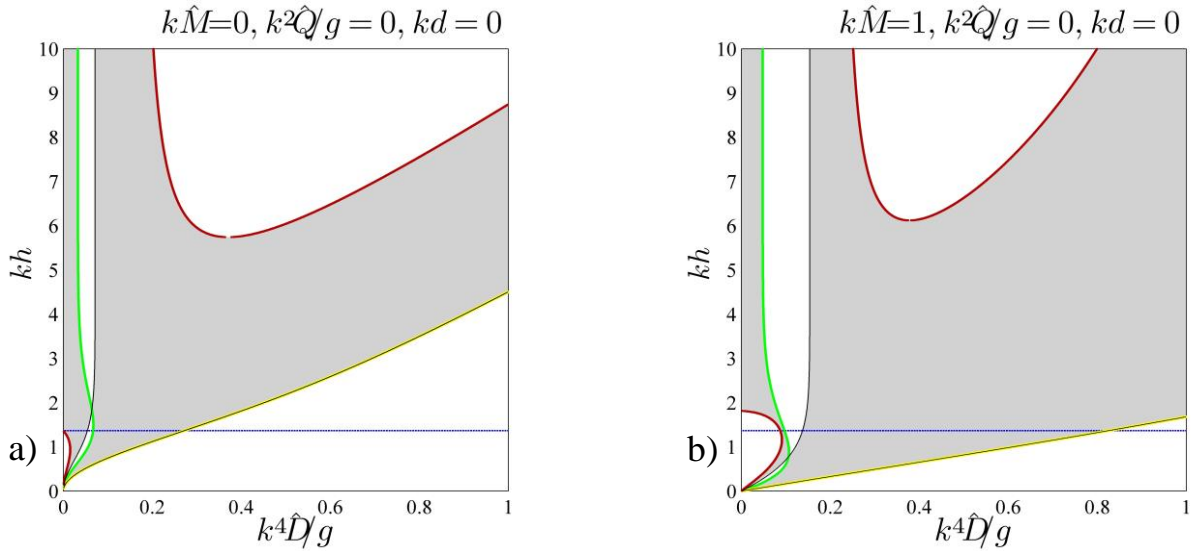

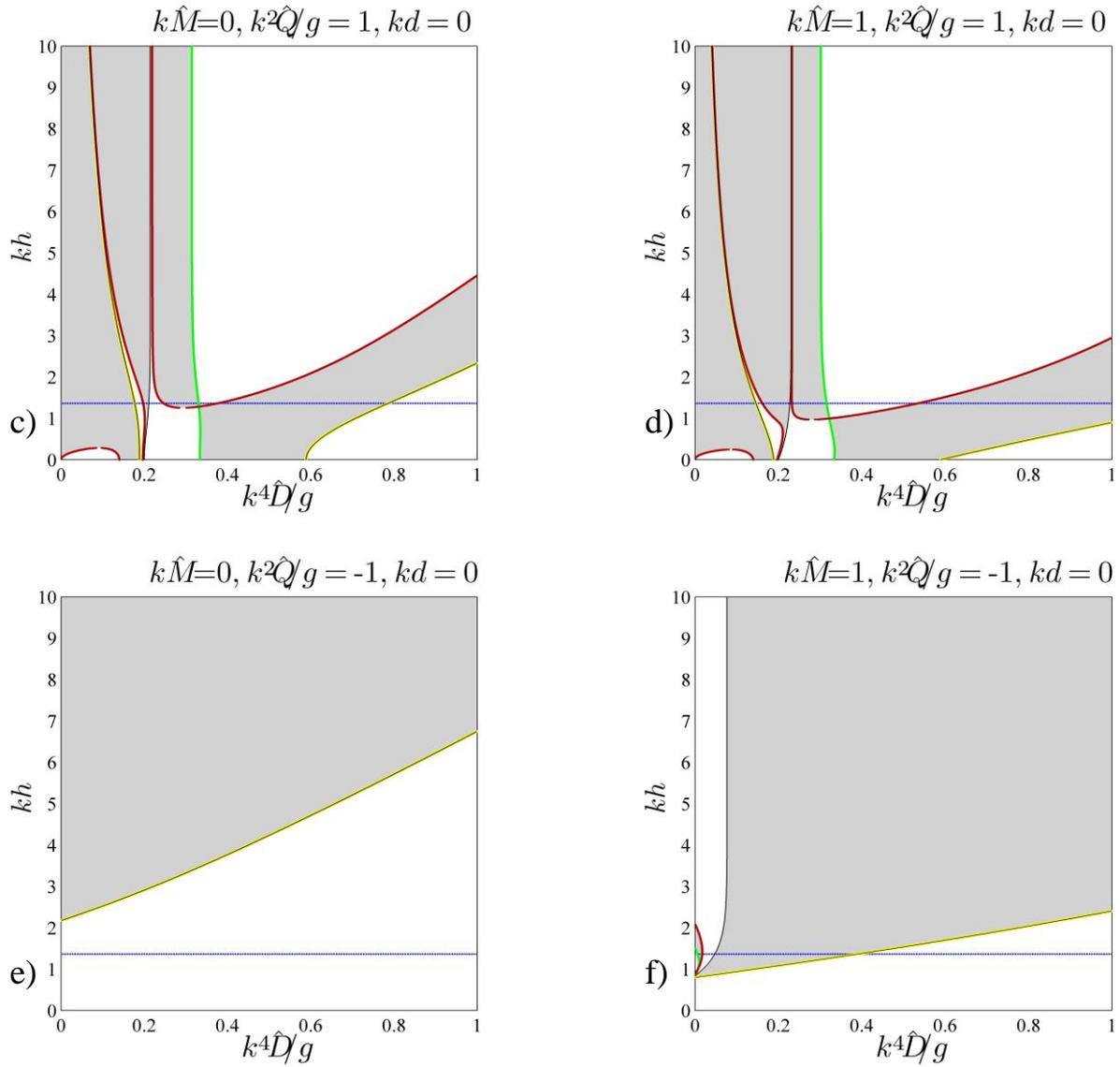

Fig. 4. (Color online.) Domains of modulational instability (shaded) for $d = 0$. Frame (a) $\hat{M} = 0$, $\hat{Q} = 0$; frame (b): $k\hat{M} = 1$, $\hat{Q} = 0$; frame (c): $\hat{M} = 0$, $k^2\hat{Q}/g = 1$; frame (d): $k\hat{M} = 1$, $k^2\hat{Q}/g = 1$; frame (e): $\hat{M} = 0$, $k^2\hat{Q}/g = -1$; frame (f); $k\hat{M} = 1$, $k^2\hat{Q}/g = -1$. The curves where coefficient $\beta$ changes its sign are shown by the light-green color. The curves where the coefficient $\alpha$ turns to zero are shown by dark-red color. Black-yellow curves correspond to the condition of the resonance with the long-scale flow, while the black curves – to the resonance with short waves. Horizontal blue dotted lines designate the condition $kh = 1.363$.

The stability property alternates each time when one of the coefficients $\alpha$ or $\beta$ changes sign. These conditions are differentiated in Figs. 3–5 by the colors of edges of the stability and instability domains. The light-green color shows the change of the sign of the dispersion coefficient $\beta$. Indeed, as follows from Fig. 2, the dispersion coefficient $\beta = -(1/2)dV/dk = -(1/2)d^2\omega/dk^2$ can change the sign in the case of flexural-gravity waves. In the range of gravity waves where the slope of the function $V(\kappa)$ is negative, the coefficient $\beta$ is positive, whereas 

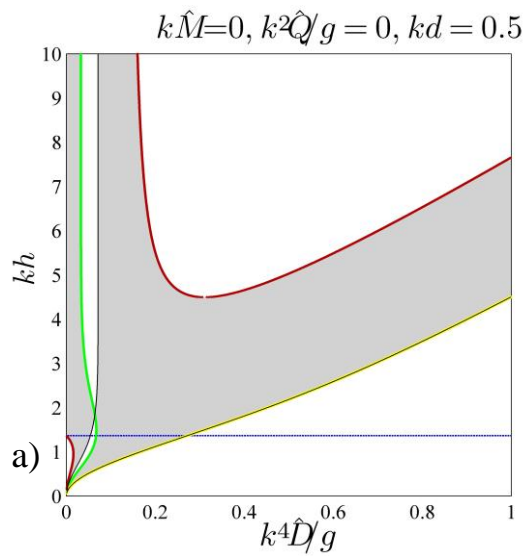
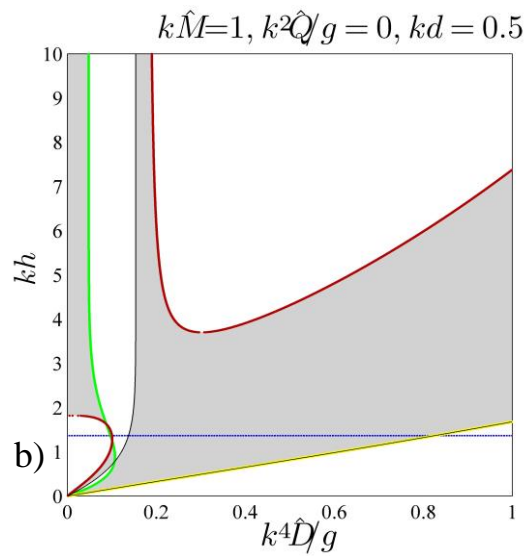
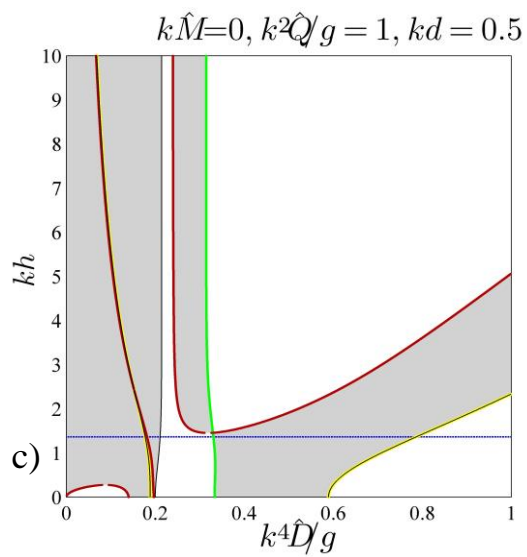
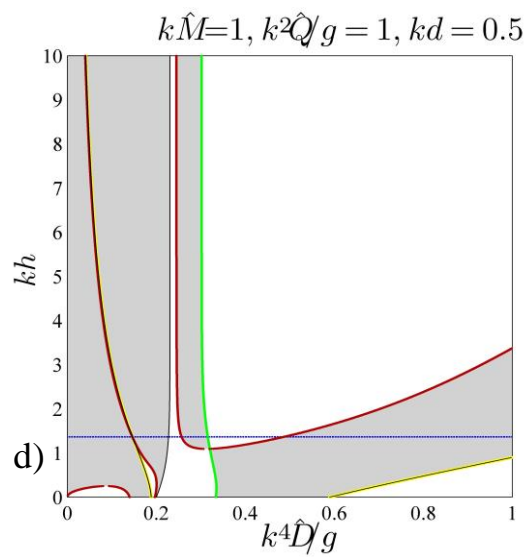
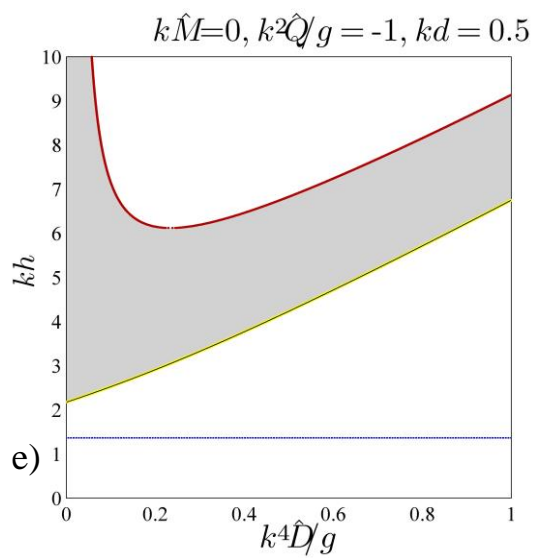
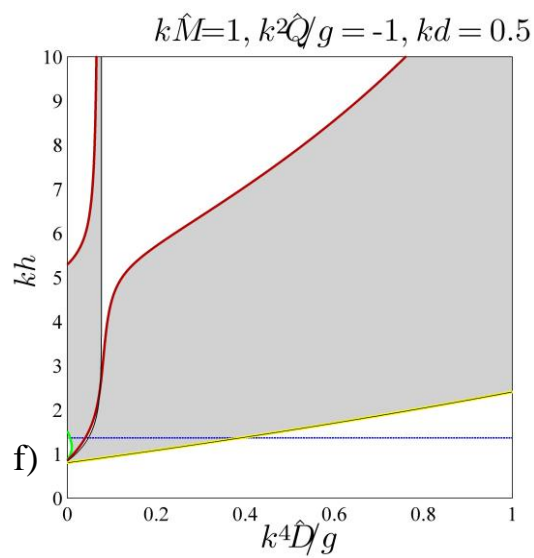

Fig. 5. (Color online) Domains of modulational instability (shaded) for $kd = 0.5$. Frame (a): $\hat{M} = 0$, $\hat{Q} = 0$; frame (b): $k\hat{M} = 1$, $\hat{Q} = 0$; frame (c): $\hat{M} = 0$, $k^2\hat{Q}/g = 1$; frame (d): $k\hat{M} = 1$, $k^2\hat{Q}/g = 1$; frame (e): $\hat{M} = 0$, $k^2\hat{Q}/g = -1$; frame (f): $k\hat{M} = 1$, $k^2\hat{Q}/g = -1$. The curves where coefficient $\beta$ changes its sign are shown by the light-green color. The curves where the coefficient $\alpha$ turns to zero are shown by dark-red color. Black-yellow curves correspond to the condition of the resonance with the long-scale flow, while the black curves – to the resonance with short waves. Horizontal blue dotted lines designate the condition $kh = 1.363$.

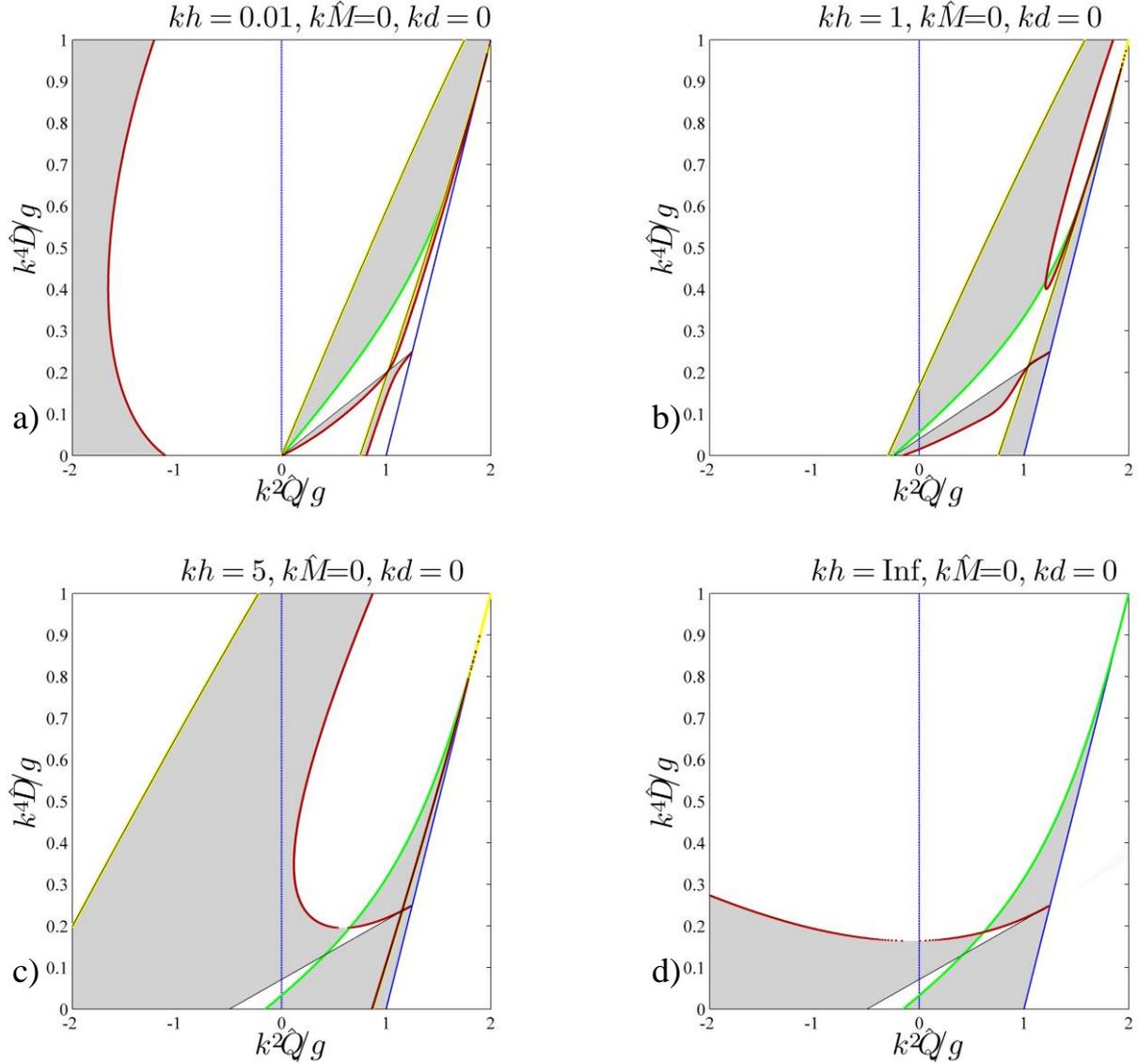

Fig. 6. (Color online.) Domains of modulational instability (shaded) in the limiting case of a thin ice sheet, $d = 0$ and $\hat{M} = 0$. Frame (a): $kh = 0.01$; frame (b): $kh = 1$; frame (c): $kh = 5$; frame (d): $kh \to \infty$. The solid blue line corresponds to the zero-frequency condition (35). The curves where coefficient $\beta$ changes its sign are shown by the light-green color. The curves where the coefficient $\alpha$ turns to zero are shown by dark-red color. Black-yellow curves correspond to the condition of the resonance with the long-scale flow, while the black curves – to the resonance with short waves. Horizontal blue dotted lines designate the condition $kh = 1.363$.

in the range of flexural waves the situation is inverse. A qualitatively similar situation takes place for gravity-capillary waves, when $M = 0$, $D = 0$ and $Q < 0$. The detailed analytic study of the dispersion is rather complicated even in the infinite-depth limit.

The nonlinear coefficient changes its sign when it passes either through the zero value (shown by red curves) or through the infinity (shown by black and black-yellow curves). Black-yellow curves correspond to the condition of the synchronism between the waves and the nonlinear induced long-scale flow,

$$gh = V^2, \tag{36}$$

which results in infinite value of the nonlinear coefficient (see Eq. (29)). It can be understood from (15) and (18) that this condition may give up to two curves for a given $kh$ in the parametric diagrams in Figs. 4 and 5.

The nonlinear resonance between the flexural-gravity wave and its second harmonic corresponds to another case when the nonlinear coefficient $\alpha$ goes to infinity (such edges are shown by black curves in the figures). This happens when the denominator in the expression for $\chi$ in Eq. (21) vanishes, $P_1 = 0$, resulting in the relation:

$$g\sigma(\sigma + 3k\hat{M}) + k^2(3 - \sigma^2)\hat{Q} - k^4(15 - \sigma^2 + 12\sigma k\hat{M})\hat{D} = 0. \tag{37}$$

This can give at most one curve for a fixed $kh$ in the parametric diagrams in Figs. 4 and 5. Note that though the denominator of the second term in the curly brackets of Eq. (26) can vanish too ($P_2 = 0$), this does not lead to a singularity of $q_1$ due to the cancellation of zero terms in the denominator and the numerator. The singularity in the coefficient $\alpha_\infty$ in Eqs. (32) remains in the limit of a deep water.

Note that the relations (36) and (37) do not depend on the ice thickness parameter $d$, therefore the corresponding boarders of instability regions in Figs. 3–5 are not affected by the change of $d$. The parameter $d$ does not enter explicitly the dispersion relation, hence the zeros of $\beta$ do not depend on $d$ directly. However, the ice thickness $d$ enters the expressions for $M$ and $D$, and hence, still influences the solution. The resulting expression for the nonlinear coefficient $\alpha$ becomes inconsistent in the formal limit when $d = 0$, but the inertia term $M$ is retained. In the scaled version of the dispersion relation (34) the ice thickness $d$ affects the solution only via $\tilde{M}$. Only zero values of the nonlinear coefficient $\alpha$ (red curves in Figs. 3–5) depend on $d$ explicitly. In practice, the parameter $kd$ is small, therefore its effect on the instability domains is not strong. Accordingly, moderate values of $M$ do not change the diagrams qualitatively either. The sets of instability diagrams for the zero ice thickness (Fig. 4) and an exaggerated value $kd = 0.5$ (Fig. 5) look generally similar. The most significant deformation of the instability diagram is observed in the case of negative $Q$, cf. Fig. 4f and Fig. 5f.

The analysis of zeros of the nonlinear coefficient $\alpha$ is fairly complicated due to the bulky terms $T_3$, $T_4$ and $T_5$ in Eq. (26). As follows from Eq. (32), in the limit of deep water the nonlinear coefficient $\alpha_\infty$ can vanish at two values of $\hat{D}$. It also follows from Figs. 4 and 5 that the number of zeros $\alpha$ can be larger in shallow water. In fact, the numerator of the coefficient $\alpha$ in the shallow-water limit contains the coefficient $\hat{D}$ in power four, hence up to four roots are possible, what agrees with the diagrams in Fig. 4 and 5.

The resonance conditions (36) and (37) are simpler for investigation. According to Figs. 4 and 5, waves can be modulationally unstable below the critical depth $kh < 1.363$ under certain combinations of the physical parameters. The instability expands down to the zero-depth limit if $Q \geq 0$ (when the ice plate is compressed). It can be readily shown that in the shallow-water limit $kh \ll 1$,

$$\omega^2 \xrightarrow[kh \to 0]{} k^2 h \left(g - k^2 \hat{Q} + k^4 \hat{D}\right), \quad V \xrightarrow[kh \to 0]{} \frac{\omega}{k} - \frac{kh}{\omega}\left(k^2 \hat{Q} - 2k^4 \hat{D}\right), \tag{38}$$

and then, the resonance condition (36) yields a quadratic equation on either $\hat{D}$ or $\hat{Q}$, which does not depend on $\hat{M}$ and $d$. The equation possesses one positive solution $\hat{D}$ for any $0 \leq k^2 \hat{Q}/g \leq 3/4$ (see black-yellow curves in Figs. 4a,b and 5a,b) and two positive roots $\hat{D}$ for $k^2 \hat{Q}/g > 3/4$ (see black-yellow curves in Figs. 4c,d and 5c,d). In particular, in the limit of dominating gravity ($k^4 \hat{D} \ll g$ and $k^2 \hat{Q} \ll g$) the resonance condition reads $\hat{Q} = 5/3 k^2 \hat{D} \geq 0$. The instability diagram for the limit of small depth is also shown in Fig. 6a; it exhibits the discussed above features.

On the other hand, as follows from the diagrams, short waves are stable with respect to self-modulation when the rigidity parameter $\hat{D}$ exceeds some value. As can be seen from Eq. (32), the group velocity in the limit $kh \gg 1$ remains finite for any finite $\hat{D}$. Therefore, for the finite coefficients of rigidity and longitudinal stress, the nonlinear resonance with long waves does not occur in the deep-water limit. This conclusion agrees with the slowly growing black-yellow edges in Figs. 4 and 5 when $\hat{D}$ increases; the corresponding curve is absent in Fig. 6d which illustrates the limit $kh \to \infty$. This also leads to the statement that the second branch of the long-wave resonance curve (the leftmost in Figs. 4c,d and 5c,d), which appears when $k^2\hat{Q}/g > 3/4$, must cross the axis $\hat{D} = 0$ at some large value of $kh$, though this point is beyond the limit of the shown graphs.

As for the super-harmonic resonance condition (37), in the shallow-water limit $kh \ll 1$, it is reduced to $\hat{Q} = 5k^2\hat{D}$ (see black line in Fig. 6a). Therefore, it always has one root when $\hat{Q} \geq 0$ and has no solutions when $\hat{Q} < 0$. In the deep-water limit $kh \gg 1$, the resonance condition (37) can be simplified too. Then, only one positive root can exist when $2k^2\hat{Q} \geq -g^2\left(1+3k\hat{M}\right)$ (see black line in Fig. 6d).

In Figs. 3–6 one can see many examples when the curves which correspond to zero and infinite values of the nonlinear coefficient $\alpha$ merge. Obviously, for the physical problem this implies the cancellation of singularities. In such situations, the NLS equation should be extended to include higher-order terms for proper description of nonlinear wave processes.

Besides the principal possibility of modulation instability, the issues on the spectral range of instability and maximum growth rate are important to be achievable under realistic conditions. It is well known [see, e.g., (Ablowitz and Segur, 1981)] that the most unstable wave within the NLS theory (30) is characterized by the perturbation wavenumber $K_m$ and the growth rate $\gamma_m$:

$$K_m = a_0\sqrt{\alpha/\beta}, \qquad \gamma_m = \alpha a_0^2, \tag{39}$$

where $a_0$ is the real wave amplitude. The range of the wavenumbers $K$ where the modulation instability occurs is $0 < K < K_m\sqrt{2}$. From the practical perspective it is constructive to express the perturbation length, $L = 2\pi/K$, and the characteristic growth time, $\tau = 2\pi/\gamma$, in terms of the wave period, $T = 2\pi/\omega$ and the wavelength, $\lambda = 2\pi/k$, respectively. Then, using Eq. (39), we obtain:

$$\frac{L}{\lambda} = n\varepsilon^{-1}, \qquad \frac{\tau}{T} = \delta\varepsilon^{-2}, \qquad n = k^2\sqrt{\frac{\beta}{\alpha}}, \qquad \delta = \frac{\omega k^2}{|\alpha|}, \tag{40}$$

where $\varepsilon = a_0 k$ is the wave steepness which is for realistic sea waves is of the order of $10^{-1}$ or less. The coefficient $n$ represents the normalized number of waves in an unstable group, and the coefficient $\delta$ is the normalized maximal growth time. For reference, these coefficients for purely gravity waves in the limit of deep water are 0.5 and 2, respectively. The quantities $n$ and $\delta$ in the domains of instability are plotted in Figs. 7 and 8 with the pseudo-color for the set of parameters like in Fig. 4. The darker color corresponds to shorter unstable wave groups (in Fig. 7) and faster modulation growth (in Fig. 8). As one can see, compared to the case of deep-water gravity waves without ice (the axis $\hat{D} = 0$ in Figs. 7a and 8a), the flexural-gravity waves can be characterized as less or more unstable depending on the ice parameters. According to Eq. (40), shorter wave groups become unstable when $\beta$ approaches zero value (inflection points of the dispersion relation), or when $\alpha$ goes to infinity. Large values of $\alpha$ also lead to faster development of the modulational instability. Interestingly, strongly unstable situations can be realized under much shallower conditions than $kh = 1.363$. As follows from Eq. (40), the conditions $\beta \approx 0$ and $\alpha \to \infty$ will result in formally very strong nonlinear self-modulation regime with small values of $n$ and $\delta$. A more accurate theory should be derived to describe these degenerate conditions.

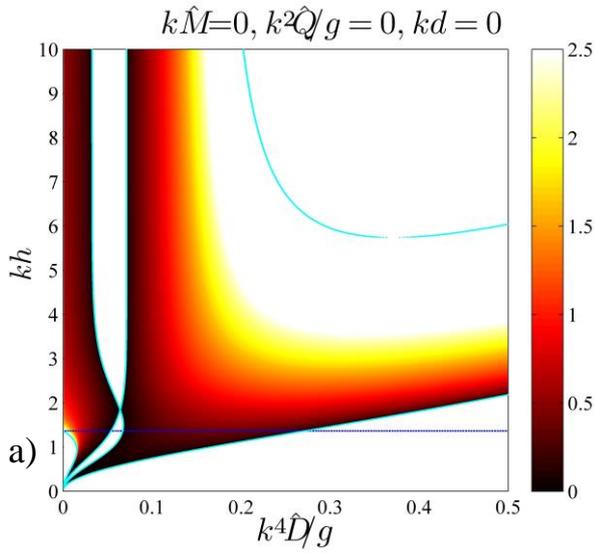
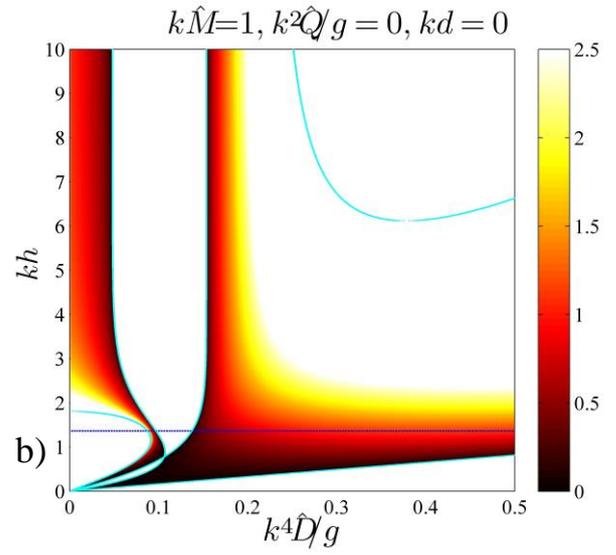
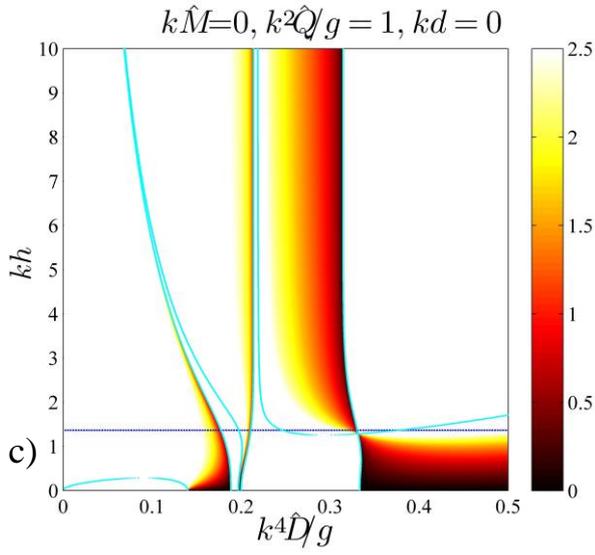
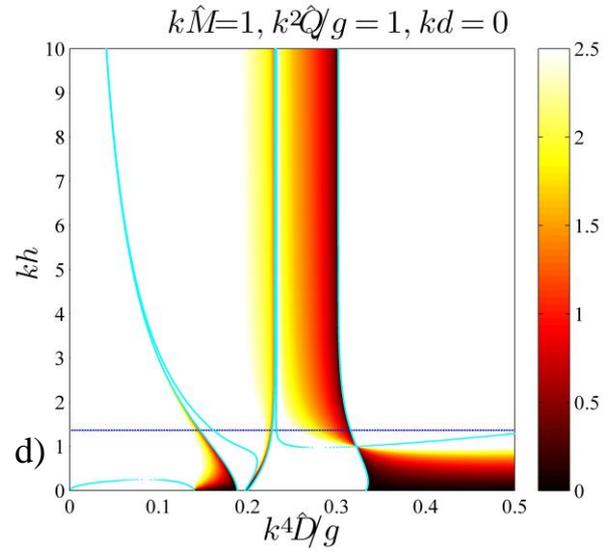
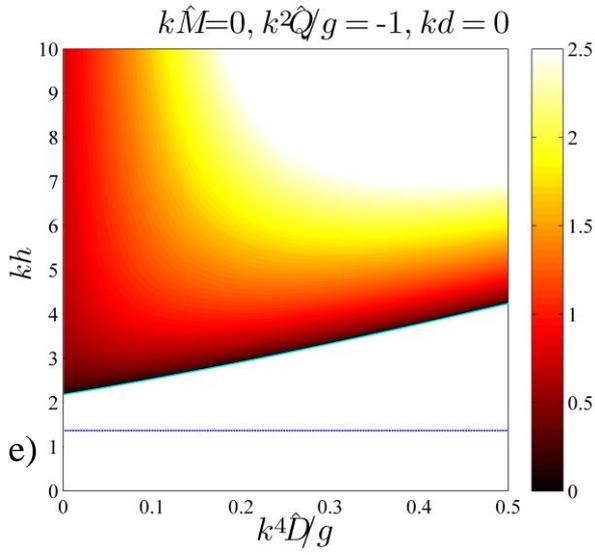
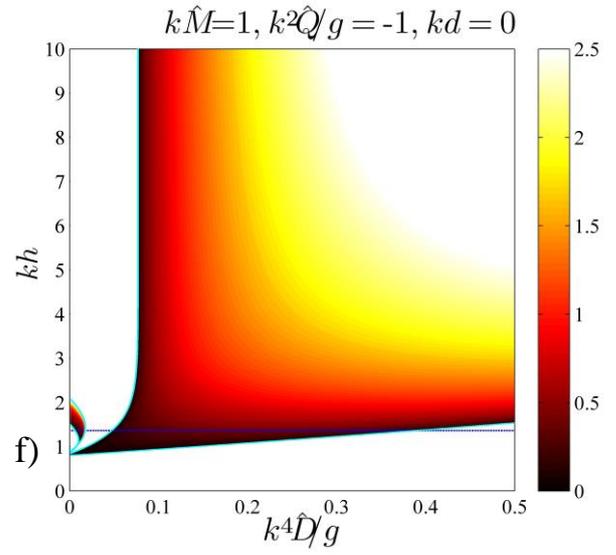

Fig. 7. (Color online.) The normalized number $n$ of waves in unstable groups are shown with the pseudo-color for $d = 0$. Frame (a) $\hat{M} = 0$, $\hat{Q} = 0$; frame (b): $k\hat{M} = 1$, $\hat{Q} = 0$; frame (c): $\hat{M} = 0$, $k^2\hat{Q}/g = 1$; frame (d): $k\hat{M} = 1$, $k^2\hat{Q}/g = 1$; frame (e): $\hat{M} = 0$, $k^2\hat{Q}/g = -1$; frame (f); $k\hat{M} = 1$, $k^2\hat{Q}/g = -1$. The edges of the instability domains are shown by cyan lines. Horizontal blue dotted lines designate the condition $kh = 1.363$. (Note the twice smaller limit of the horizontal axis here compared to Fig. 4).

To gain some insight how the considered effects can manifest in realistic conditions, we present in Fig. 9 the parameter of growth time $\delta$ in the plane of the ice thickness $d$ and carrier wavelength $\lambda = 2\pi/k_0$ for a few values of depth $h$ and compression parameter $\tilde{Q} = Q/\sqrt{\rho g D}$. The diagrams for $n$ look qualitatively similar. The horizontal dotted lines in each frame show the boundary $k_0 h = 1.363$. It should be beard in mind that the genuine scaled time of the modulational growth is $\delta \varepsilon^{-2}$ provided that $\varepsilon = 0.1$ or less.

## 6. Discussion and Conclusion

In this paper, we have studied the modulation property of flexural-gravity waves on a water surface covered by a compressed heavy ice sheet of a given thickness. We have derived the nonlinear Schrödinger equation with coefficients that depend on the ice parameters and water depth (they are given in the electronic supplement). The new theory is consistent with the earlier findings by Ablowitz and Segur (1979, 1981), though does not agree with the result by Liu and Mollo-Christensen (1988). Conditions, when a quasi-sinusoidal wave becomes unstable with respect to the amplitude modulation, have been investigated.

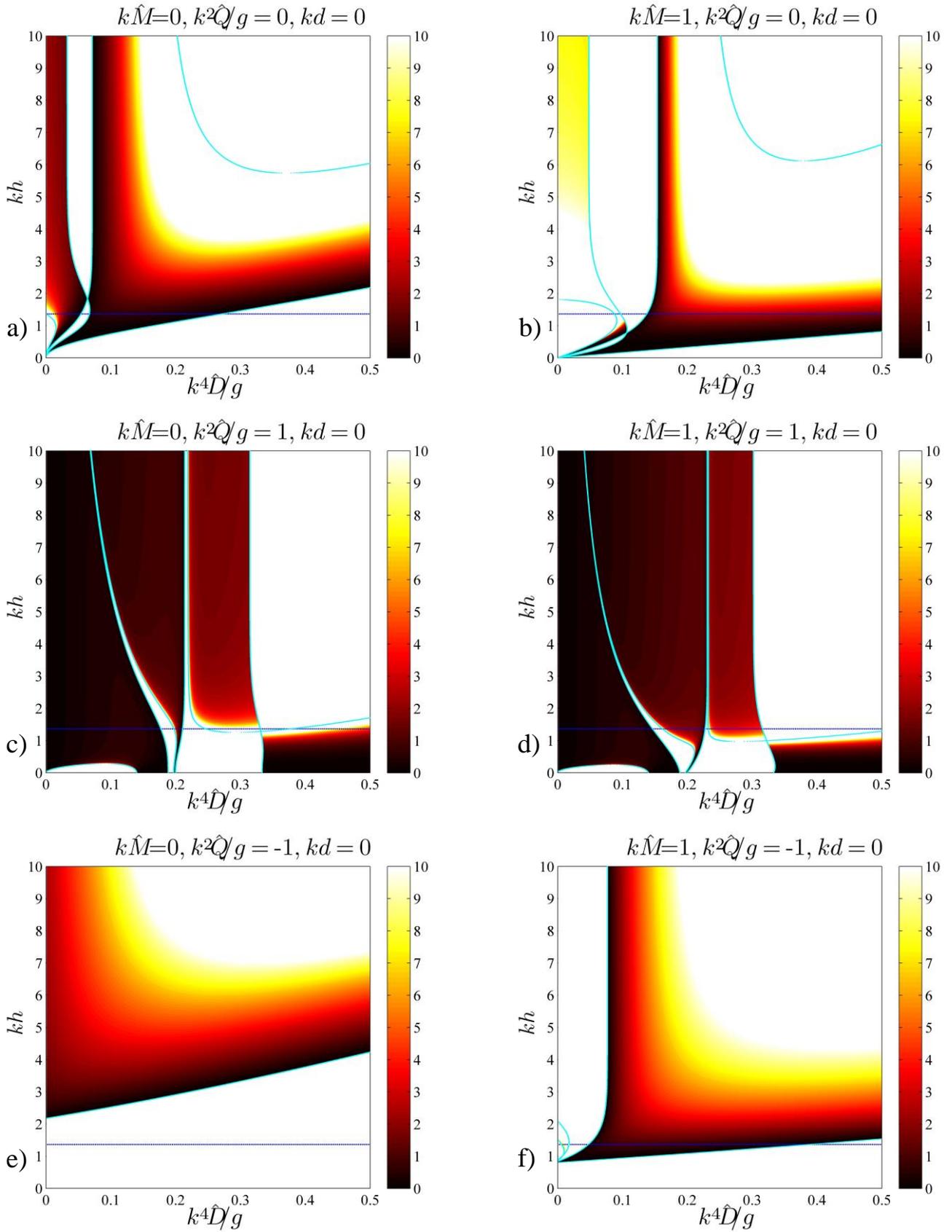

Fig. 8. (Color online.) The normalized maximum growth times $\delta$ are shown with the pseudo-color for $kd = 0.5$. Frame (a) $\hat{M} = 0$, $\hat{Q} = 0$; frame (b): $k\hat{M} = 1$, $\hat{Q} = 0$; frame (c): $\hat{M} = 0$,

$k^2 \hat{Q}/g = 1$; frame (d): $k\hat{M} = 1$, $k^2\hat{Q}/g = 1$; frame (e): $\hat{M} = 0$, $k^2\hat{Q}/g = -1$; frame (f); $k\hat{M} = 1$, $k^2\hat{Q}/g = -1$. The edges of the instability domains are shown by cyan lines. Horizontal blue dotted lines designate the condition $kh = 1.363$. (Note the twice smaller limit of the horizontal axis here compared to Fig. 4).

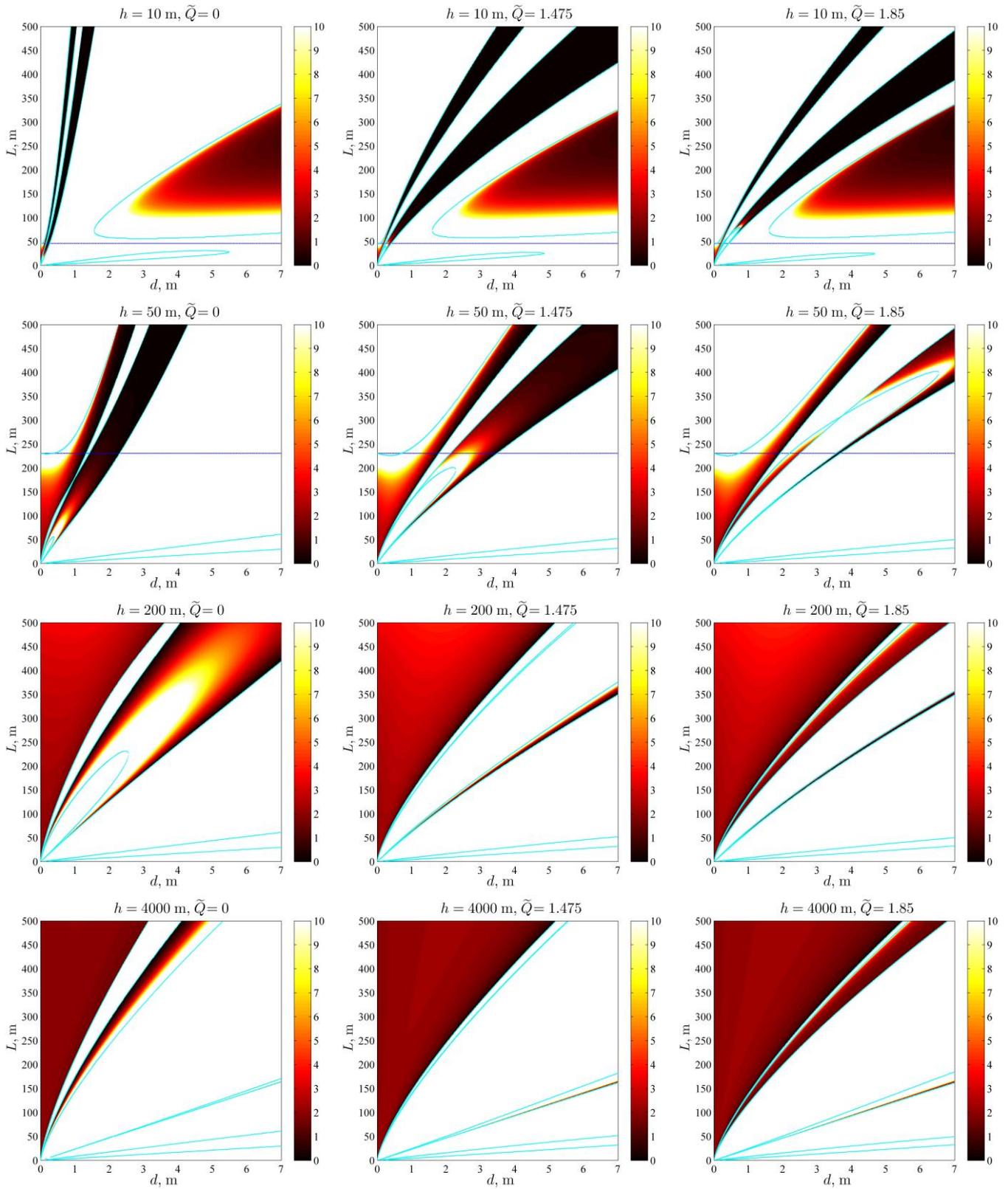

Fig. 9. (Color online.) The normalized maximum growth times $\delta$ are shown with the pseudo-color in the parameter plane of ice thickness $d$ and carrier wavelength $L$ for the different values of compression parameter $\tilde{Q}$. The edges of the instability domains are shown by the cyan curves. Left column – $\tilde{Q}=0$; middle column – $\tilde{Q}=1.475$; and right column – $\tilde{Q}=1.85$. In the upper row the water depth is $h = 10$ m; in the second row $h = 50$ m; in the third row $h = 200$ m; and in the last row $h = 4\,000$ m (see the legends above the frames).

It is well known that the modulational instability in open oceans is the most efficient when the ocean depth is infinite. Our analysis reveals that the presence of an ice cover not just allows the development of the modulational instability, but for some combinations of the ice and depth parameters can lead to even stronger nonlinear self-modulation in different senses. This pertains to the condition on the minimum length of unstable wave groups; the maximum growth rate; the threshold depth when the instability occurs. As one can see from Figs. 3–6, domains of the modulation stability and instability are bizarrely interspersed depending on combinations of the physical parameters.

The modulational instability is related to the dangerous rogue wave phenomenon when abnormally large waves can emerge from rather regular initial perturbations (Onorato et al., 2001; Kharif et al., 2009). When such rogue waves occur on the top of an ice sheet, they can lead to ice destruction; this problem deserves further study. In the context of large floating artificial constructions, the obtained stability-instability diagrams may help to design safer constructions – by choosing the structure characteristics which correspond to stable wave regimes. The importance of highlighting instability domains is in the understanding when bright and dark solitary waves can exist (sketchy examples of such solitary waves are given in Fig. 10). The former can arise in the process of development of the modulational instability, whereas the latter can appear in the modulationally stable regions on the parameter plane. Solitons are long-lived coherent wave patterns that possess their own specific dynamical features. For example, envelope solitons demonstrate different from linear waves amplitude growth rates when transform adiabatically in inhomogeneous conditions or due to pumping (Onorato and Proment, 2012; Slunyaev et al., 2015); also, they can form bound states with repeated extreme wave occurrence (Ducrozet et al, 2021), and so on. When bright solitons interact with significant background waves, they may be described by breather solutions of the nonlinear Schrodinger equation, which are the mathematical prototypes of oceanic rogue waves. The presence of soliton- or breather-type waves alters the wave statistics so that it can remarkably deviate from the Gaussian statistics (Onorato et al., 2013; Randoux et al., 2016; Slunyaev and Kokorina, 2017).

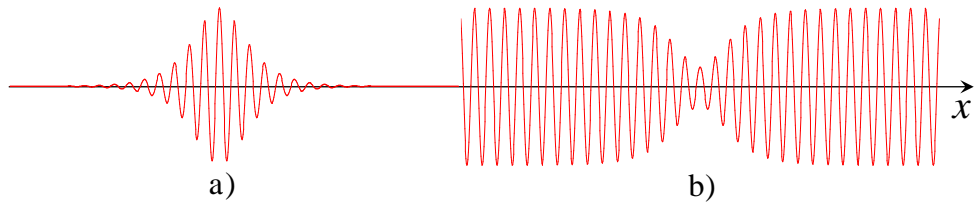

Fig. 10. (Color online.) Examples of a bright soliton (a) and a dark soliton (b).

Nonlinear envelope solitary waves on the surface of finite-depth water covered by ice were considered in the recent publication by Il'ichev (2021). For the particular case of the water depth $h = 55$ m and the ice sheet thickness $d = 1$ m, the author derived from the fully nonlinear Euler equations a solution for the envelope soliton with the carrier wavelength that corresponds to the minimum of a phase speed in the fluid of a finite depth. In such a case, the phase ($C = \omega/k$) and group ($V = d\omega/dk$) speeds become equal. It was shown that the theoretically derived shape of the envelope solitary wave with the wavenumber $k \approx 0.045$ m$^{-1}$ ($\lambda \approx 138$ m) and amplitude $a_0 = 2$ m was very similar to that which follows from the NLS equation. Therefore, with such parameters and wave steepness $ka_0 = 0.09$, the NLS equation well describes flexural-gravity waves in water covered by ice. One can expect applicability of the NLS equation to this kind of wave until the steepness remains small $ka_0 \ll 1$.

Apparently, bright solitons can emerge in the wakes behind moving pressure sources such as landing or taken-off aircrafts, snowmobiles, skaters on frozen rivers or lakes, etc. A similar problem was considered by Berger and Milewski (2000) who discovered the formation of gravity-capillary lumps (fully localized in space two-dimensional solitons) in the wake behind a moving source on a thin water layer. The problem of generation of NLS bright solitons by moving loads on the floating ice plate or by topographic effects due to a flow around underwater hills is very topical. Such problems are more complex but solvable; they yet wait for their solution.

Note that the cubic NLS equation degenerates or becomes invalid, and higher-order theories should be developed when some of the nonlinear or dispersion coefficients in this equation either vanish or become singular. These specific cases correspond to boundaries between the domains of modulation stability and instability and yield the maximum unstable growth rates which may be significantly reduced within the more accurate examination. As known, the wave attenuation due to dissipative effects can stabilize the modulational instability as well (Segur et al., 2005). A reliable theoretical model of flexural-gravity wave attenuation is not developed yet. According to experimental data (Squire, 2020), the decay rate of wave amplitude is a power-type function of frequency, $\omega^p$, where the exponent $p$ varies in the range 1.9–3.6 depending on the ice property. A well-known fact is that the wave motion from the Southern Ocean can penetrate the Antarctic marginal ice zone up to 400 km (Squire, 2020). This suggests that the wave attenuation under certain conditions can be relatively weak. In the first approximation of wave amplitude, the dissipation effect can be taken into account by introducing a linear dissipation term in the NLS equation (Alberello and Părău, 2022); this leads to the downshifting of the spectral peak and a less than exponential energy decay. Our estimates show that the dissipative term is weak compared to the nonlinear and dispersive terms in the NLS equation of the paper (Alberello & Părău, 2022) model with the exponent $n = 3$ if the phase speed of FGWs is relatively small, $V_p \ll (\rho g^2 A^2/\rho_l d \nu_w)^{1/3}$, where $A$ is the wave amplitude, and $\nu_w$ is the water viscosity. Using the ice-water parameters after Eq. (34) and setting $A = 1$ m, $\nu_w = 10^2$ m$^2$/s, we obtain $V_p \ll 1$ m/s. Such values are quite realistic in the vicinity of a phase speed minimum, especially in the compressed ice (we did not present here the plot of the phase speed for FGWs but it is qualitatively similar to the plot of the group speed shown in Fig. 2). From the formal point of view, wave self-modulation remains significant if the dissipation term is of the order of $\varepsilon^2$ or smaller.

Consideration of the problem beyond the one-dimensional geometry should open even more intriguing perspectives. Under certain physical conditions, one may expect such situations when a wave collapse becomes possible [see, for example, (Marchenko and Shrira, 1992)]; this can lead, apparently, to ice cover buckling and crash in finite time. All aforementioned unsolved problems can be a challenge for future studies.

**Acknowledgements**

The authors are thankful to M. Klein for the valuable discussion. A.S. acknowledges the support from the International Visitor Program SMRI of the University of Sydney and is grateful to the School of Mathematics, Physics and Computing of the University of Southern Queensland, Australia for the hospitality. The part of research reported in Sections 3 and 5 was supported by the Russian Science Foundation (grant No. 22-17-00153). The remaining parts were funded by the Ministry of Science and Higher Education of the Russian Federation (grant No. FSWE-2021-0009) and by the Council of the grants of the President of the Russian Federation for the state support of Leading Scientific Schools of the Russian Federation (grant No. NSH-70.2022.1.5).